# Emergence of Techno-social Norms in Cognitive Radio Environments

Ligia Cremene, *Member, IEEE,* D. Dumitrescu

*Abstract*—The aim of this paper is to explore the potential of Game Theory (GT) in extracting rules of behaviour for emerging Cognitive Radio environments. We revisit the commons approach to unlicensed spectrum and try to show that a commons can be basically regulated from the inside out. GT simulations of CR interactions reveal the emergence of certain equilibria mirroring behaviours/trends?. Once these ?trends identified, norms may be expressed and then embedded into machines (CRs). Internalized norms may thus become the alternative to external enforcement of rules. We call these emerging norms *techno-social norms (TSNs)*. TSNs could eventually become a means of regulating the use of unlicensed spectrum and making open spectrum access feasible. Open spectrum access scenarios are considered and analysis is performed based on reformulations of two game theoretical models: Cournot and Bertrand. The standard oligopoly models are reformulated in terms of radio resource access in unlicensed bands. In order to capture the large variety of CR interaction situations, several GT equilibrium concepts are considered: Nash, Pareto, Berge-Zhukovskii, and Lorenz. In order to capture the heterogeneity of CR interactions, the standard GT model is enriched allowing players to be biased toward different types of equilibrium (or rationality). An evolutionary game-equilibrium detection method is used. Numerical simulations bring relevant insights on the problem of autonomy vs. regulation in emerging CR environments. Relying on extensive GT simulations, some rules of behaviour – to be expanded into techno-social norms – may be derived.

## I. INTRODUCTION

Current spectrum regimes are based on a highly prescriptive approach, centralized control and decisions [1], [2], [3]. However, regulatory bodies have acknowledged the inefficiencies inherent in command-and-control spectrum regulation. More flexible and market-driven approaches have been considered [4], [5]. Coase's vision [6] of private rights in spectrum has been challenged by the idea that new technologies make any control of spectrum unnecessary [7], [8], [9]. Although the administrative approach makes it easier for the regulators to ensure avoidance of excessive interference, to tailor appropriate license conditions based on guard bands and maximum power transmission levels [1], [2], [5], traditional spectrum planning has been proved to be valid only for a certain generation of technology. It is a slow process that cannot keep up with innovations and new technologies [1], [2], [4].

Once of the consequences of spectrum planning anachronism is the large amount of underused spectrum, perceived as spectrum scarcity. Studies have shown that up to 90% of the radio spectrum remains idle in any one geographical location [1], [5], [11], [12]. New spectrum bands are being released around the world (e.g. 2.6 GHz in Europe, the 800 MHz digital dividend, 700 MHz and AWS – 1700/2100 MHz in the U.S., etc.). To add to the picture, existing spectrum bands are being deregulated to allow coexistence of 2G, 3G, and 4G technologies.

Cognitive radio technology is seen as the key enabler for next generation communication networks, which will be spectrum-aware, DSA networks [11], [12], [27]. In this paper the autonomy vs. regulation problem in spectrum access [12] is addressed from a game theoretical perspective. We try to answer the questions: *Is there a common-sense knowledge that can be detected and established in an emerging Cognitive Radio environment? Can this be embedded as rules of behaviour in CRs so that to reduce the need for external enforcement of rules?* To what extent can rules of behaviour be expressed so that they may be internalized by intelligent communicating machines like CRs.

Our focus is unlicensed spectrum. Unlicensed does not mean unregulated [1]. The problem with free access is that there is always the risk that it will eventually lead to interference and over-saturation, i.e., the 'tragedy of the commons' or the 'spectrum tragedy' [13].

Our approach points to a built-in normative system where the unlicensed spectrum is managed as a regulated commons [3]. This leads to a discussion about the nature of a commons. In the context of unlicensed spectrum usage, the commons concept needs to be revisited. The concept of common-pool resources (CPRs) as defined by Ostrom [14] does not cover entirely the nature of the unlicensed spectrum as a resource. A

Manuscript received January 15, 2012. This work was supported in part by CNCSIS –UEFISCDI, Romania, PD, project number 637/2010.

L. C. Cremene is with the Technical University of Cluj-Napoca, Romania (e-mail: Ligia.Cremene@com.utcluj.ro).

D. Dumitrescu, is with Babes-Bolyai University, Cluj-Napoca, Romania. (corresponding author, phone: 0040742182359; fax: 0040264401917, e-mail: ddumitr@cs.ubbcluj.ro).

The authors are also with the Romanian Institute of Science and Technology.

more nuanced and appropriate concept is needed.

Ostrom uses the term common-pool resources to refer to "resource systems regardless of the property rights involved [15]. CPRs include natural and human constructed resources in which *(i)* exclusion of beneficiaries through physical and institutional means is especially costly, and *(ii)* exploitation by one user reduces resource availability for others [15]. These two characteristics – difficulty of exclusion and subtractability – create potential CPR dilemmas in which people following their own short-term interests produce outcomes that are not in anyone's long-term interest" [15].

When we discuss spectrum access, the dilemmas are no longer about the long-term interests of the users but, as the effects of interference are immediate, they are about the diminishing loss and harm produced to others.

Our approach relies on looking at the unlicensed spectrum as a special type of common resource to be shared among very dynamic users. Social norms often serve to regulate the use of a common resource. In our opinion, functioning of the social norms, in general, is based on identity. Identity is the one that keeps social norms from breaking down. Both individual and group identity is built in time, and social norms arise by repeated interactions [16], [17]. But, in unlicensed spectrum we have anonymous users and fluid groups. There are no guarantees that they are likely to interact with one another on a regular basis [16]. Therefore identity seems to play no role here.

We consider that the standard conditions [14], [17] for norm emergence by repeated interactions are not fulfilled in CR environments. A CR built-in normative system may be more suitable.

In the proposed approach, CR interactions are seen as strategic interactions [23], [ ]: the utility of one CR depends on the actions of all the other CRs in the environment. An intuitive metaphor describing a CR environment may be that of a society. The CRs' strategic interactions may be seen as a kind of social interactions. Emergent norms in such systems may be called *techno-social norms*. This could be a first step towards a new paradigm of technical systems including interactions, norms, and social values.

In this context we propose the use of extensive game theoretical simulations of CR interactions as a means of extracting norms for emerging CR environments. The extracted norms may than be programmed/embedded into CRs as operational rules.

We see *techno-social norms* as a means of regulating the use of unlicensed spectrum and making open spectrum access feasible.

Embedding techno-social norms, and eventually values – like fairness – into intelligent machines takes interactions to another level.

Moreover, the proposed approach may be useful per se. CRs, conceived as intelligent agents, may directly interact based on strategic interactions. In this situation game equilibria may describe a certain kind of desirable situation from which no agent has any incentive to deviate. Several equilibrium types are considered. Some point to selfish interactions, some are capturing more equitable solutions.

The paper is structured as follows: Section II presents our proposed GT approach to CR emerging environments. Section III provides some basic insights to non-cooperative game equilibria detection. The reformulation of Cournot and Bertrand game theoretic models in terms of open spectrum access scenarios is described in Section IV. Section V presents and discusses the numerical results obtained from simulations. The conclusions are presented in Section VI

## II. GAME THEORETICAL APPROACH TO COGNITIVE RADIO EMERGING ENVIROMENTS

Cognitive radio technology is seen as the key enabler for next generation communication networks, which will be spectrum-aware, DSA networks [11], [12], [27]. In a CR environment users strategically compete for spectrum resources in dynamic scenarios.

Radio resource allocation and dynamic spectrum access may be described as games [18], [19], [20], [21], [22]. A game is any situation in which cognitive entities/agents/players interact. The problem in Game Theory is that players do not normally know in advance what strategy the other players will choose. That is why game equilibria, representing standard solution concepts, are important. Several types of non-cooperative game equilibria are described in Section III.

The most frequently used steady-state solution concept is the Nash Equilibrium (NE) [18], [23], [24], [25]. Yet, there are other equilibria that may be relevant for spectrum access scenarios and especially for extracting rules of behaviour.

In order to illustrate the proposed approach we consider two oligopoly competition models – Cournot and Bertrand – reformulated in terms of radio resource access. Simultaneous access situations are considered and modelled as one-shot games. Continuous and discrete forms of the games are analyzed. We assume the wireless nodes are equipped with cognitive radios (i.e. have perfect channel sensing and RF reconfiguration capabilities) [2], [3], [12].

We analyze different types of game equilibria, as they describe various types of strategic interactions between cognitive agents – each CR's action directly affects the others' payoffs.

Papers proposing game theoretical approaches to dynamic spectrum access [3], [18], [22], etc. discuss the Nash equilibrium and Pareto optimality (efficiency). NE [25] may be seen as an 'as good as it gets' situation, but it suffers from excessive competition among self-regarding players and the outcome of the game is not always Pareto efficient [26].

Assuming there is no reason for the players to be other then selfish, the following questions guide our approach:

*Q1: Can we go beyond the Nash equilibrium in one-shot resource access games?*

*Q2: How can we get to a desired equilibrium in a highly dynamic CR emerging environment?*

*Q3: Can we get to a solution where each player gets what it needs, the solution is both optimal and fair?*

Starting with these in mind, we consider other types of equilibria, besides Nash and Pareto, namely: *(i)* the *Lorenz equilibrium* [28], [29], [30] – usually not a GT but a Multi Criteria Optimization solution concept – which captures a fairness equilibrium in a spectrum sharing situation, *(ii)* the *Berge-Zhukovskii equilibrium* [31], [32], [33] – capturing a limit situation, where unreliable resources are involved, and *(iii)* the *joint Nash-Pareto and Pareto-Nash equilibria* [24] – capturing the heterogeneity of the players, considered as biased towards different types of rationality.

## III. NON-COOPERATIVE GAME EQUILIBRIA

The following equilibria are considered: Nash, Pareto, Nash-Pareto, Berge-Zhukovskii, and Lorenz.

A strategic-form game model has three major components: a finite set of players, a set of actions, and a payoff/utility function which measures the outcome for each player determined by the actions of all players [26].

Based on these three components, a game may be defined as a system $G = (N, (S_i, u_i), i = 1, …, n)$ where:

(i) $N$ represents the set of $n$ players, $N = \{1, …, n\}$.

(ii) for each player $i \in N$, $S_i$ represents the set of actions $S_i = \{x_{i1}, x_{i2}, …, x_{im}\}$; $S = S_1 \times S_2 \times … \times S_N$ is the set of all possible game situations;

(iii) for each player $i \in N$, $u_i : S \to R$ represents the payoff function.

A strategy profile (strategy or action vector) is a vector $x = (x_1, ..., x_n)$, where $x_i \in S_i$ is a strategy (or action) of player $i$. By $(x_i, x^*_{-i})$ we denote the strategy profile obtained from $x^*$ by replacing the strategy of player $i$ with $x_i$, i.e. $(x_i, x^*_{-i}) = (x^*_1, x^*_2, ..., x^*_{i-1}, x_i, x^*_{i+1}, ..., x^*_n)$.

Let $I$ be a set (coalition) of agents. $(x_I, x^*_{-I})$ denotes the strategy profile in which $i \in I$ chooses the individual strategy $x_i$, and each $j \in N - I$ chooses $x^*_j$.

*Nash equilibrium*

A Nash equilibrium is a strategy profile in which each player's strategy is a best reply to the strategies of the other players [10], [25]. Hence, at NE, no player can improve her payoff by unilateral deviation [23], [24].

*Pareto equilibrium*

Considering two strategy profiles $x$ and $y$ from $S$, the strategy profile $x$ is said to Pareto dominate the strategy profile $y$ (and we write $x < P\ y$) if the payoff of each player using strategy $x$ is greater or equal to the payoff associated to strategy $y$ and at least one payoff is strictly greater. The set of all non-dominated strategies (Pareto frontier) represents the set of Pareto equilibria of the game [24].

The *Berge-Zhukovskii equilibrium* [31] was introduced as a solution for games that do not have a Nash equilibrium, or have more than one. The Berge-Zhukovskii equilibrium (B-Z) touches cooperative issues and leads to the idea that cooperation may be brought in a non-cooperative framework.

A strategy $x^*$ is a Berge-Zhukovskii equilibrium, if, when at least one of the players of the coalition $N-\{i\}$ deviates from her equilibrium strategy, the payoff of player $i$ for the resulting strategy profile would be at most equal to her payoff $u_i(x^*)$ for the equilibrium strategy.

Otherwise stated, the payoff of player $i$ decreases if one or more of the other players deviate from the B-Z equilibrium [32].

Formally, a strategy profile $x^* \in S$ is Berge-Zhukovskii equilibrium if the inequality
$$u_i(x^*) \geq u_i(x^*_i, x_{N-i})$$
holds for each player $i = 1, ..., n$, and $x_{N-i} \in S_{N-i}$.

At Berge-Zhukovskii equilibrium, each player maximizes the payoff of the other players. It reflects an altruistic tendency or reciprocation behaviour [33]. Indeed, according to the above inequality, when any player $i \in I$ plays her strategy $x^*_i$, from the Berge equilibrium $x^*$, she obtains a maximum payoff only when the remaining players $-i$ play the strategy $x_{N-i}$ from the Berge equilibrium $x^*$.

*Lorenz equilibrium*

The most popular solution concepts in game theory, Nash and Pareto equilibrium, have some limitations when applied to real world problems. Nash equilibrium rarely ensures maximal payoff and the Pareto equilibrium is a set of solutions that is often too hard to process. However, there is an equilibrium concept that provides a small set of efficient solutions and is equitable for all players – the Lorenz equilibrium (LE).

The Lorenz dominance relation [28], [29], [30] also called equitable dominance relationship, is considered. Crowding differential evolution may be used to detect the Lorenz-optimal solutions [35].

Lorenz dominance, as a refinement of Pareto dominance, deals with fair optimization problems. In addition to the initial objective aiming at maximizing individual utilities, fairness refers to the idea of favouring well-balanced utility profiles.

*Joint Nash-Pareto equilibrium*

In an $n$-player game consider that each player $i$ acts based on a certain type of rationality $r_i$, $i = 1, …, n$. We may consider a two-player game where $r_1 = Nash$ and $r_2 = Pareto$. The first player is biased towards the Nash equilibrium and the other one is Pareto-biased. Thus, a new type of equilibrium, called the joint Nash-Pareto equilibrium (N-P), may be considered [24]. The recently introduced Nash-Pareto equilibrium concept [24] captures a game situation where players are biased towards different types of rationality: Nash and Pareto. Their actions are biased towards different equilibria and the resulting equilibrium has hybrid characteristics.

*Evolutionary equilibrium detection*

An evolutionary technique for equilibria detection, based on appropriate generative relations [24] that allow the comparison of strategies, is considered.

An appealing technique is the use of generative relations and evolutionary algorithms for detecting equilibrium strategies. The payoff of each player is treated as an objective and the generative relation induces an appropriate dominance concept, which is used for fitness assignment purpose.

Evolutionary multiobjective algorithms are thus suitable tools in searching for game equilibria [24], [30], [33], [34].

Numerical experiments aim the detection of pure equilibria or a combination of equilibria, paralleling cognitive radios' interaction. An adaptation of the popular NSGA2 [34] has been considered. Similar results are obtained when using a multiobjective method based on the Crowding Differential Evolution algorithm [35] or on the Topological Species Conservation Algorithm [36].

All the considered equilibria are computationally stable with respect to the evolutionary detection technique.

A population of strategies is evolved. A chromosome is an $n$-dimensional vector representing a strategy profile $x \in S$. The initial population is randomly generated. The population model is generational. The population of strategy profiles at iteration $t$ may be regarded as the current equilibrium approximation. Subsequent application of the search operators is guided by a specific selection operator induced by the generative relation. Successive populations produce new approximations of the equilibrium front, which hopefully are better than the previous ones.

## IV. SPECTRUM ACCESS SCENARIOS. NUMERICAL EXPERIMENTS

In order to assess open access scenarios, two oligopoly game models are considered and reformulated in terms of radio access: Cournot and Bertrand [23]. The difference between the two models lies in the player's strategy – quantity for Cournot (number of accessed channels) and price for Bertrand (number of non-interfered symbols per channel). The cognitive radios may use different strategies: the number of accessed channels or the number of non-interfered symbols.

In the Cournot economic competition players independently and simultaneously choose quantities of a product to be sold. In the Bertrand oligopoly players simultaneously choose prices for the products they sell [23].

In order to illustrate open spectrum access situations, scenarios with two CRs simultaneously trying to access the same resource are considered. The resource is considered either a set of available channels or an available capacity.

For the sake of clarity, CR strategies and payoffs are represented two-dimensionally. Many-player scenarios have also been analyzed. The results represent a sub-set of more extensive simulations.

For equilibria detection the evolutionary technique from [24] is considered. A population of 100 strategies has been evolved using a rank based fitness assignment technique. In all experiments the process converges in less than 50 generations. Our tests show that the evolutionary method for equilibrium detection is scalable with respect to the number of available channels [37].

### A. Open spectrum access – reformulation of Cournot model

We consider an open access spectrum access scenario that can be modelled as a reformulation of the Cournot oligopoly game [18], [23], [38].

Suppose there are $n$ radios attempting to access the same whitespace simultaneously. Each radio $i$ may decide the number $c_i \in [0, \infty)$ of simultaneous channels to access. Because the number of non-interfered symbols depends on the total number of accessed channels, each CR's action directly affects the others' payoffs.

The question is how many simultaneous channels should each CR access in order to maximize its operation efficiency?

Based on the above scenario, a reformulation of the Cournot game may be as follows:

*Players* cognitive radios attempting to access a certain set of channels $W$;

*Actions* the strategy of each player $i$ is the number $c_i$ of simultaneous accessed channels;
A strategy profile is a vector $c = (c_1,…,c_n)$.

*Payoffs* the difference between a function of goodput $P(c)c_i$ and the cost $Kc_i$ of simultaneously accessing $c_i$ channels.

We consider a linear inverse demand function in which the number of non-interfered symbols $P(c)$ is determined from the total number $c_i$ of accessed channels (occupied bandwidth).

The demand function can be defined as:

$$D(C) = \begin{cases} W - C, & if\, C < W, \\ 0, & otherwise, \end{cases}$$

where $W > 0$ is the parameter of the inverse demand function and $C = \sum_{i=1}^{n} c_i$, is the aggregate number of accessed channels.

The goodput for CR $i$ is $P(c)c_i$. Radio $i$'s cost for supporting $c_i$ simultaneous channels is $C_i(c_i)$.

The payoff of CR $i$ may then be written [18] as:

$$u_i(c) = P(c)c_i - C_i(c_i).$$

In general, $P$ decreases with the total number of implemented channels and $C_i$ increases with $c_i$ (more bandwidth implies more processing resources and more power consumption) [18]. If these effects are approximated as linear functions, the payoff function can be rewritten as

$$u_i(c) = \left(W - \sum_{k=1}^{n} c_k\right)c_i - Kc_i,$$

where
$W$ is the set of available channels (whitespace), and
$K$ is the cost of accessing one channel.

The Nash equilibrium, considered as the solution of this game, can be calculated as follows:

$$c_i^* = (W - K)/(n+1), \forall i \in N.$$

*Reformulated Cournot – numerical experiments*

Simultaneous access scenario simulation results are presented for the Cournot competition. Two cognitive radios simultaneously try to access a set of available channels. Continuous and discrete strategies are considered.

*Continuous instance Cournot game*

The considered whitespace size is $W = 24$ (24 available channels). The cost of accessing one channel is $K = 3$. The action space is continuous.

The emerging behaviour of the radio environment is captured by the detected equilibria (Fig. 1): Nash, Pareto, Nash-Pareto, Pareto-Nash, Berge-Zhukovskii, and Lorenz. The six types of equilibria are obtained in separate runs.

In the continuous case, a NE to the game exists and is unique (7,7). The NE corresponds to the scenario where each of the two CRs activates 7 channels (from 24 available).

The Pareto equilibrium (Fig. 1) describes a situation where the number of active channels for each CR is no more than half of the available set (lies in the range [0, 10.5]). Moreover, the sum of active channels at Pareto equilibrium is less than the sum of active channels at NE. The sum of simultaneously accessed channels is maximum at NE. Therefore, we may consider that this situation indicates an efficient use of the available spectrum, in terms of occupancy and fairness.

Fig. 2 illustrates the payoffs of the two players, $u_1(c_1, c_2)$ and $u_2(c_1, c_2)$, for the Cournot resource access game. The NE indicates the maximum number of channels a Nash-biased CR may access without decreasing its payoff (Fig. 2).

Lorenz equilibrium lies at the middle of the Pareto front both in the strategy space (5.25, 5.25) (Fig. 1) and in the payoff space (55.13, 55.13) (Fig. 2). This equilibrium is especially relevant for selecting a fair/equitable Nash equilibrium among multiple ones, as it will be the case with discrete strategies. The NE closest to LE is the most suitable candidate. We may notice that, at Lorenz equilibrium, the number of accessed channels for each CR is lower than at NE, yet the payoffs are higher.

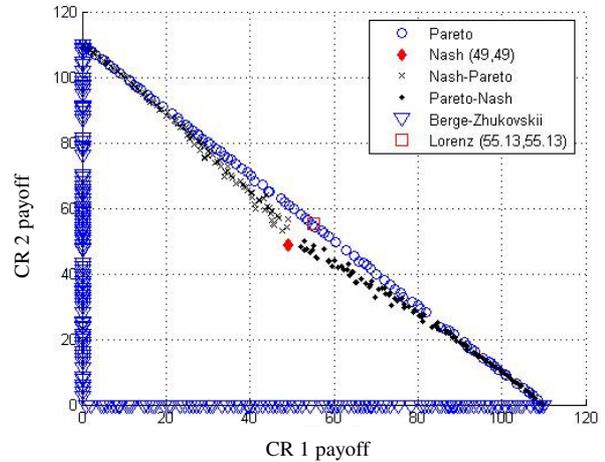

Fig. 2. Cournot payoffs of the evolutionary detected equilibria: Nash (49, 49), Pareto, Lorenz (55.13, 55.13), Berge-Zhukovskii, N-P, and P-N. Continuous Cournot modelling – two cognitive radios ($W = 24$, $K = 3$).

The set of Pareto-optimal solutions is rather large, whereas the Lorenz equilibrium consists of a single solution - the one that provides the maximum possible payoff for all players. Similar to the Lorenz equilibrium, the Nash equilibrium also consists of a single solution that is also a fair one (equal number of channels, equal payoffs), but compared to the Lorenz solution it provides a lower payoff for both players (Fig.2).

The joint Nash-Pareto equilibrium is achieved for the strategies on the N-P front (Fig. 1). We may notice that, in some cases, the Nash-Pareto strategy enables the CR to access more channels than the NE strategy (7,7). In the performed experiments the P-N equilibrium is symmetric to the N-P equilibrium. It is interesting to notice that none of the N-P or P-N strategies actually reach NE.

The payoffs for the Pareto strategies (Fig. 2) are in the range [0, 110] and their sum is always larger than for the NE payoff (49,49). For each strategy of the Nash-Pareto equilibrium the Pareto-player has a higher payoff. The Nash-player payoff is smaller in a Nash-Pareto situation than in a case where all the players are Nash-biased.

Even if the N-P strategies allow the CRs to access more channels, the payoffs are smaller than for the Pareto strategies. This may be explained by the presence of interference increasing with the number of accessed channels.

In general, NE suffers from excessive competition among self-regarding players and the outcome of the game is not always Pareto efficient [26]. In the Cournot case, the NE is close to the Pareto front (Fig. 2) which indicates that even if the users are self-regarding the outcome of a simultaneous access may be satisfactory.

B-Z equilibria are represented (Fig. 1) by strategies of the form *(a,0)* or *(0,b)*, $a, b \in [0, 10]$. This indicates that there are situations where each player gets to maximize the other's payoff. This also means that if one player selects a number of channels, the other selects none. According to B-Z semantics, if the idle user decides to access one or more channels, the active CR's payoff decreases. This may describe a limit situation where resources are unreliable or very unstable, and may illustrate a coexistence problem. The B-Z maximum payoff achievable by one CR is equal to that at Pareto equilibrium (110.25).

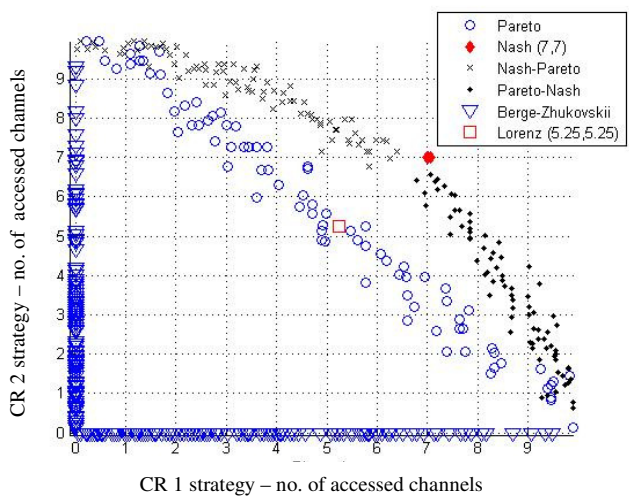

Fig. 1. Cournot strategies - Evolutionary detected equilibria: Nash (7,7), Pareto, Lorenz (5.25, 5.25), Berge-Zhukovskii, Nash-Pareto, and Pareto-Nash. Continuous Cournot modelling – two cognitive radios ($W = 24$, $K = 3$).

*Discrete instance Cournot game*

Cournot typically allows continuous transitions between player actions. However, as in our modelling the actions concern the number of accessed channels, we may consider a discrete version of the game; the model becomes thus more realistic. For the discrete case, the equilibria cannot be detected by standard calculus procedures, yet the chosen evolutionary detection method [24], [34] proves to be efficient for discrete games also.

Fig. 3 illustrates the evolutionary detected strategies for the discrete-form Cournot resource access game (W=24, K=3). Their corresponding payoffs are captured in Fig. 4.

continuous-form game – there are two extra Nash equilibria: (6,8) and (8,6) (Fig. 3). This means that each CR has three strategies that ensure maintaining its payoff, provided it does not deviate from these strategies. The extra NEa also represent a N-P and P-N strategy, respectively.

However, the extra NEa are no longer perfectly equitable (equal number of channels, equal payoffs) as in the continuous-form game. This is where Lorenz equilibrium helps in choosing one of the Nash equilibria, namely the closest to LE, the most fair one. This turns out to be exactly the continuous-form NE (7,7) (Fig . 1).

A rule of behaviour that may be extracted from the reformulation of Cournot competition is the following: *if the CR wants to avoid the decrease of its payoff it should not attempt to access more channels than the number indicated by the NE strategy* (7 channels for the current situation W=24, K=3) (answer to *Q2*).

Also, if a CR wants to go beyond the NE it may choose the LE which is both equitable and profitable (answer to *Q1*). In this case each CR accesses fewer channels than at NE (5 instead of 7) and achieves a higher payoff (55 instead of 49).

### B. Open spectrum access – reformulation of Bertrand model

In the Bertrand economic competition, producers compete by varying the product price and thus adjusting the demand. A constant unit cost and linear demand function are assumed. Players decide their actions independently and simultaneously. This time, their strategy is the price instead of the quantity. The model was extensively used for pricing problems, including spectrum trading [3], [39].

The player's strategy in a Bertrand competition is the price. Considering a reformulation of the game in spectrum access terms, the equivalent of the price $P(c)$ may be each CR's target number of non-interfered symbols per channel. This further translates into capacity demand, $D(p)$.

The Bertrand competition for spectrum resource access may be reformulated as follows:

*Players* — the cognitive radios attempting to transmit a number of symbols (competing for amounts of an available capacity $W$).
*Actions* — the strategy of each CR $i$ is a target number $p_i$ of non-interfered symbols per channel;
*Payoffs* — the difference between a function of goodput $p_i D(p_i)$ and the cost $C_i(D(p_i))$ of transmitting $N$ symbols.

Let us consider $n$ cognitive radios competing for a given available throughput $W$. The objective of each CR is to activate a subset $c_i$ of channels in order to satisfy its current demand level (i.e. target throughput). We can thus write:

$$W = \sum_i c_i N_c,$$

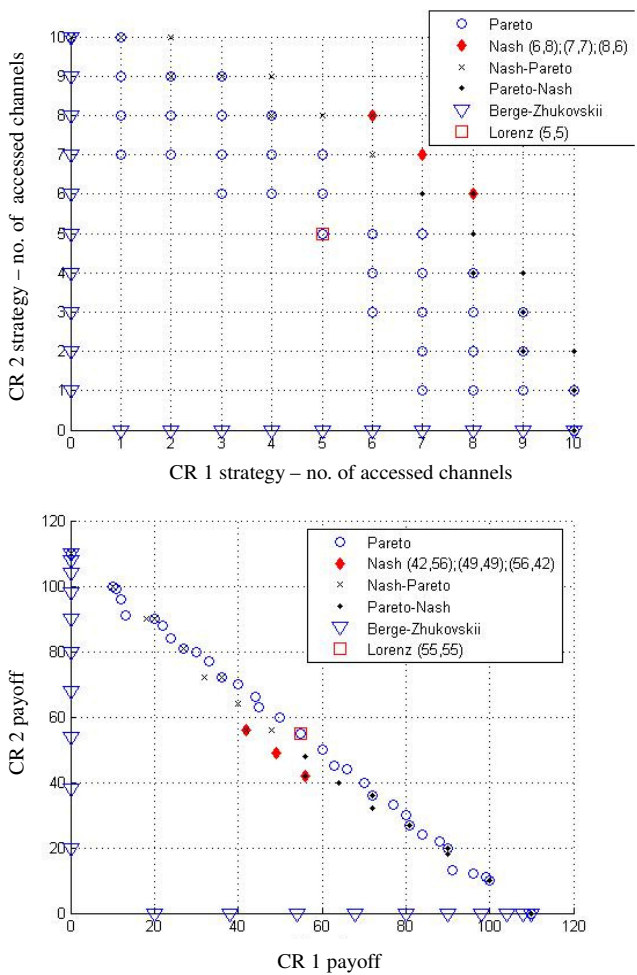

Fig. 4. Discrete Cournot modelling. Payoffs of the evolutionary detected equilibria: three Nash eq. (42,56); (49, 49); (56,42), Pareto eq., N-P, P-N, Berge-Zhukovskii, and Lorenz (55,55).

We notice the existence of multiple Nash equilibria, in this case – three. Besides (7,7) – which is the NE of the where $N_c$ is the total number of symbols/channel.

The demand function of accessing a channel is a decreasing function $D$ of $p$, where $p$ is the number of non-interfered symbols.

The payoff of player *i* is defined as
$$u_i(p_i) = p_i D_i(p_i) - C_i(D_i(p_i)), i = 1,2.$$
where $C_i$ is a cost function.

Let us assume that the demand function is defined as
$$D(p) = W - p \text{ for } p \leq W \text{ and}$$
$$D(p) = 0 \text{ for } p > W.$$

The demand *D* becomes zero when the requested throughput *p* reaches the available capacity *W*.

The payoff function of CR *i* can be expressed as:

$$u_i(p_1, p_2) = (p_i - K)(W - p_i), p_i < p_j$$
$$= 1/2(p_i - K)(W - p_i), p_i = p_j$$
$$= 0, p_i > p_j.$$

when $K < W$ – that is the overhead cost *K* of accessing a channel is lower then the available capacity *W*.

*Reformulated Bertrand – numerical experiments*

Open spectrum access simulation results are presented for the Bertrand competition. Two cognitive radios simultaneously try to access amounts of an available capacity *W*. Continuous and discrete strategies are considered.

*Continuous instance Bertrand game*

Fig. 5 and Fig. 6 qualitatively illustrate the equilibrium situations for two CRs simultaneously trying to access a limited available capacity *W*. The strategy of each CR is the number of non-interfered symbols per channel, $p_i$, that they request.

In a scares resource situation, the lower the CR's target of non-interfered symbols per channel is, the higher the chances are for the CR to get access to one or several channels. On the other hand, as the number *P(c)* of non-interfered symbols per channel decreases, the need for channels (the demand) increases. Thus, a CR willing to maximize its goodput will attempt to occupy as many low-rate channels as possible.

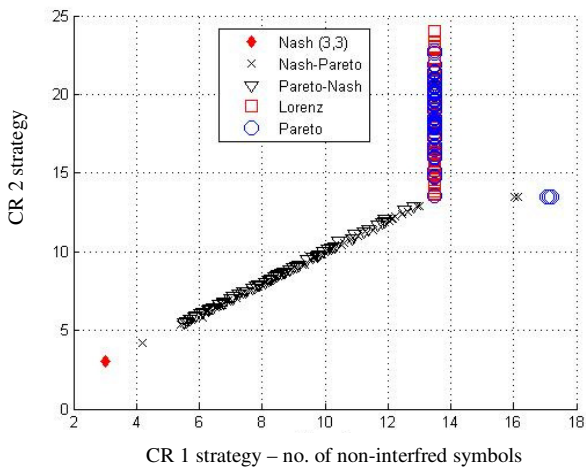

Fig. 5. Bertrand strategies - Evolutionary detected equilibria: Nash (3, 3), Pareto, N-P, P-N, and Lorenz. Continuous-form Bertrand modelling – two cognitive radios (W=24, K=3).

for one CR at a time (Fig. 6). This may indicate that some sort of scheduling or sequential access scheme would be necessary.

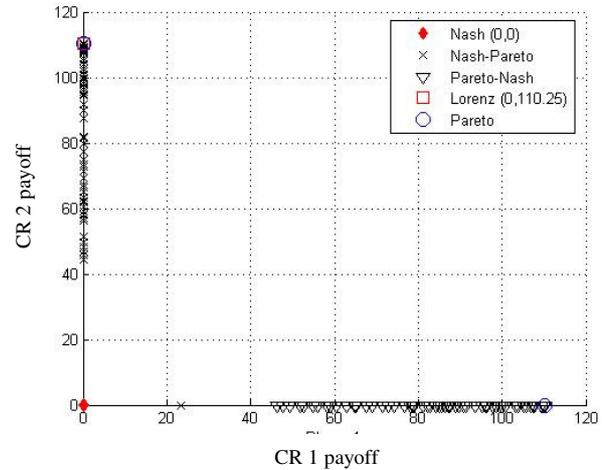

Fig. 6. Bertrand payoffs of the evolutionary detected equilibria: Nash (0, 0), Pareto (110.25, 0), (0, 110.25), N-P, P-N, and Lorenz (0,110.25). Continuous Bertrand modelling – two cognitive radios (W=24, K=3).

The joint strategies, Nash-Pareto and Pareto-Nash, which are symmetric in terms of strategies and payoffs (Fig. 5, Fig. 6), do not ensure a least equitable distribution of payoffs: the Pareto player is favoured. In the continuous form of this game model, the Nash player's payoff is zero.

The experiment results suggest using Bertrand model to describe 'win-lose' situations, when considering one or more Pareto players. The NE describes an equal, zero-payoff, situation.

*Discrete instance Bertrand game*

Discrete-form Cournot and Bertrand oligopoly games exhibit multiple Nash equilibria. Multiple equilibrium points to choose from give more flexibility in obtaining a fair and efficient resource allocation.

Discrete Bertrand modelling reveals the possibility of non-zero payoffs for both players: Nash, Pareto, and Lorenz equilibria (Fig. 8).

The evolutionary detected equilibria and payoffs are captured in Fig. 7 and Fig. 8 (W=24, K=3).

There are two Nash equilibria: (3,3), (4,4). This time there is a non-zero NE payoff : (10,10). One of the NEa, namely (4,4), is also a P-N strategy.

We may also notice (Fig.8) that, in the discrete game, a Pareto intermediate situation (55,55) appears in the payoff space. In this point boths CRs have non-zero payoffs. This turns out to be also the Lorenz equilibrium. Answer to Q3

As the utility of Lorenz equilibrium is sensibly higher than that of Nash equilibrium, there are reasons to go beyond Nash equilibrium. (answer to Q1)

N-P and P-N equilibria do not involve equal individual strategies and the payoffs are unbalanced (Fig. 8), reflecting a 'win-lose' situation (which is typical for the standard, continuous Bertrand, cf. Fig. 6).

A rule of behaviour that may be extracted is: in order to get non-zero payoffs, in a Bertrand competition, the fairness issue cannot be ignored. Like in many social situations, fairness is profitable. The number of non-interfered symbols for LE is higher than for NE and it ensures a higher payoff.

## V. CONCLUSIONS

This paper introduces the concept of techno-social norms for emerging CR environments. These norms can be derived from extensive GT simulations of strategic interactions between CRs. In order to investigate the relevance of certain game equilibrium concepts for the problem of open spectrum access, Cournot and Bertrand oligopoly game models are reformulated in terms of spectrum resource access. Continuous and discrete strategies are considered.

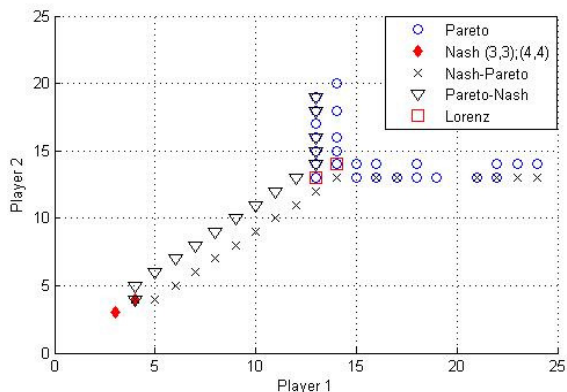

Fig. 7. Bertrand discrete strategies - Evolutionary detected equilibria: Nash ((3,3), (4,4), Pareto, N-P, P-N, and Lorenz (13,13), (14,14); (W=24, K=3).

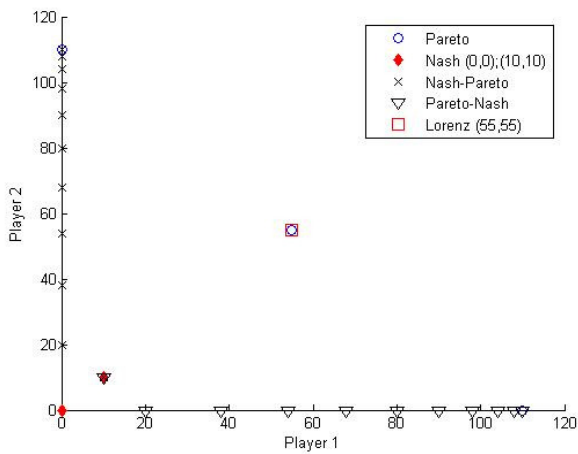

Fig. 8. Bertrand discrete payoffs of the evolutionary detected equilibria: Nash (0, 0), (10,10), Pareto (110,0), (0, 110), (55,55), N-P, and P-N and Lorenz (55,55). (W=24, K=3).

Several types of game equilibria are analyzed, describing various types of strategic interactions between CRs. Besides the standard Nash equilibrium we interpret the Pareto, Lorenz, and Berge-Zhukovkii equilibria. Joint Nash-Pareto and Pareto-Nash equilibria are also considered. Discrete instances of oligopoly games exhibit multiple Nash equilibria. Discrete equilibria reveal new, more profitable and equitable situations. Fairness is profitable… The existence of multiple equilibrium points gives more flexibility in obtaining a fair and efficient resource allocation. As Lorenz equilibrium is the most equitable and profitable for both players it may serve for selecting a Nash equilibrium. In the discrete instance of the Bertrand competition, fairness also proves profitable – equitability here brings non-zero payoffs for both players.

The Berge-Zhukovskii equilibrium describes a borderline situation where any non-zero choice of the idle radio implies decreasing the payoff of the active CR.

The considered games and equilibria enable the description and anticipation of a large variety of possible situations. Relying on such experiments, some rules of behaviour, to be expanded into techno-social norms may be derived. This will eventually lead to internalizing values like fairness into machines.